# Improving magnetic nanothermometry accuracy through mixing-frequency excitation

Silin Guo,[1] Jay Liu,[3] ZhongZhou Du,[1] and Wenzhong Liu[1,2,a)]

[1] School of Artificial Intelligence and Automation, Huazhong University of Science and Technology, Wuhan 430074, China
[2] Key Laboratory of Image Processing and Intelligent Control, Huazhong University of Science and Technology, Wuhan 430074, China
[3] Ningbo Chuanshanjia Electrical and Mechanical Co., Ltd., Ningbo 315400, China

In this study, we proposed a temperature model of magnetic nanoparticle relaxation and a phase measurement method under a mixing-frequency excitation field, which can improve the temperature accuracy of magnetic nanothermometry. According to the Debye-based magnetization model for magnetic nanoparticles, the phases at the mixing frequencies are used to solve the relaxation phase delay to the magnetic field with the higher frequency. The method could improve the signal-to-noise ratio of the magnetic response signal, and also weaken the phase shift of the detection coils caused by temperature changes. Experimental results show that the method can achieve static temperature measurement error less than 0.1K and dynamic temperature measurement error less than 0.2K.

## I. INTRODUCTION

Metabolism is the set of life-sustaining chemical processes that enables organisms transform the chemical energy stored in molecules into energy that can be used for cellular processes. As heat is a byproduct of metabolism, temperature variation at the cell level happens all the time due to cellular activities. This means that when the temperature of the human body is abnormal, it is likely that the internal physiological conditions deviate from the normal state[1-3]. Consequently, temperature sensing of living cells would provide not only insight into a variety of cell events, but also a grasp of cellular pathological state, permitting the development of diagnostic and therapeutic techniques for some diseases[4]. However, temperature mapping of living cells in situ requires the accuracy of a thermometer better than 0.1K[5]. Non-invasive and accurate temperature measurement is a key challenge in medical temperature imaging.

Magnetic methods for measuring temperature are some of the most promising in vivo temperature imaging methods because they allow measurements to be made within living organisms[6]. Among published magnetic measurement methods, the low-frequency ones with the field frequency lower than 1kHz are based on the Langevin equation[7-13], and the middle-frequency ones in the range from 1kHz to 100kHz are based on the Debye model[14-16].

It is generally believed that increasing the frequency of the AC excitation magnetic field can increase the response speed and the imaging speed of temperature measurement[14, 17, 18]. However, the accurate measurement of relaxation time at high frequency actually faces the bottleneck of signal-to-noise ratio, which affects the temperature measurement error, as shown in Fig.FIG. 1. Due to the relaxation mechanism in magnetic nanoparticles, the increase of magnetic field frequency would cause energy loss, which leads to the attenuation of harmonic amplitude and affects the signal-to-noise ratio of the magnetic response signal[19-21], as shown in Fig.FIG. 2(a). In contrast, increasing the magnetic field strength can improve the signal-to-noise ratio, but shorten the effective relaxation time[18, 22], as shown in Fig.FIG. 2(b). As it is difficult to consider both frequency and intensity by using a single-frequency magnetic excitation field, the temperature errors of published methods have difficulties in achieving stable resolutions better than 0.3K[15].

Except the excitation field, the measurement system also affects the results of temperature measurement. At present, a pair of coils with the same parameters are generally used in the detective system to eliminate the remanence[23]. As it is almost impossible for two coils to be exactly the same in practical application, the coil properties affected by temperature would change the amplitude and phase of the detection signal, and then influence the temperature accuracy.

---

a) Author to whom correspondence should be addressed. Electronic mail: lwz7410@hust.edu.cn.

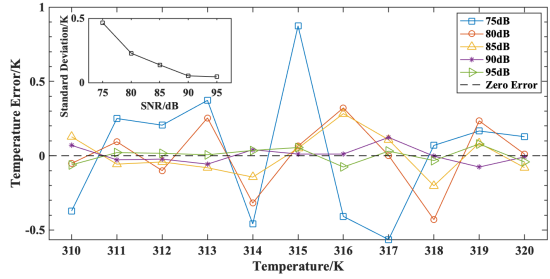

FIG. 1 Simulation of temperature error under different SNR. Assume $D_H = 30$nm, $\eta = 10^{-3}$Pa·s, $H = 1.5$ mT and $f_H = 5$kHz.

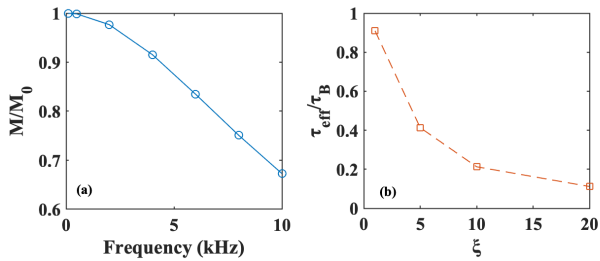

FIG. 2 The effect of excitation magnetic field on the magnetic response of magnetic nanoparticles. (a) The relationship between the magnetic response intensity and the magnetic field frequency. (b) The relationship between the effective relaxation time and the magnetic field intensity. $\xi = \mu_0 m_s H / k_B T$ is related to the field intensity.

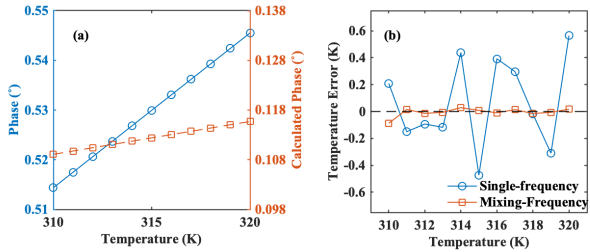

FIG. 3 Simulations of (a) the phases at high frequency $f_H$ and (b) temperature errors of the single-frequency model and mixing-frequency model. Assume $D_H = 30$nm, $f_H = 6$kHz, $f_L = 1.5$kHz, the simulation results of phase changed with temperature is based on $\Phi(T) = \tan^{-1} L(T)/R(T)$.

For this reason, we have studied the mixing-frequency harmonic under two sinusoidal magnetic fields applied simultaneously. As shown in Fig.FIG. 3(a), compared with the directly measured phase, the phase at high frequency calculated by the mix-frequency phases basically independent of temperature. As shown in Fig.FIG. 3(b), the temperature error is markedly lower using mixing-frequency excitation than single-frequency excitation for the same power. The maximum temperature errors for mixing-frequency and single-frequency excitation magnetic fields are 0.08K and 0.56 K, respectively.

In this paper, we herein propose two method to improve magnetic nanothermometry based on the relaxation time of magnetic nanoparticle under the mixing-frequency field, which is the sum of a magnetic field with the higher frequency and a filed with the lower frequency. The experimental results show that calculating the phase at the higher frequency by the phases at the mixing frequencies can avoid the decay of the signal at the higher frequency, and suppress the influence of coil mismatch and temperature change on the phase measurement. In the temperature range of 310-320K for biomedical applications[24-27], the method can achieve static temperature measurement error less than 0.1K and dynamic temperature measurement error less than 0.2K.

## III. EXPERIMENT MODEL AND METHOD

### A. Model of Relaxation time under Mixing-frequency Field

Two sinusoidal magnetic fields are applied simultaneously as follow: one with the lower frequency $f_L$ and amplitude $H_L$, written as $H_L\cos(2\pi f_L t)$; the other with the higher frequency $f_H$ and amplitude $H_H$, written as $H_H\cos(2\pi f_H t)$ [28, 29, 28, 29]. The two frequencies of the two magnetic field are both lower than 100kHz. The excitation field is the sum of these two fields, which can be described as the following:

$$H(t) = H_L\cos(2\pi f_L t) + H_H\cos(2\pi f_H t) \quad (1)$$

Under the AC excitation field, MNPs have two relaxation mechanisms: the physical rotation of particle in viscous medium is called Brownian relaxation, and magnetic dipole flipping inside a stationary particle is called Néel relaxation. The total relaxation process is a parallel model of the two relaxation schemes[19]. Shown as Fig.FIG. 4, Brownian relaxation would dominate when the diameter of particles is larger than 20nm,

$$\tau_{eff} \approx \tau_B = \frac{3\eta V_H}{k_B T} \quad (2)$$

, where $\eta$ is the viscosity of the matrix fluid and $V_H$ is the hydrodynamic volume of MNP. As the excitation frequency increases, the phase delay φ produced by the relaxation processes have a relationship with relaxation time τ, which can be described by Debye model[19]:

$$\chi(\omega) = \frac{\chi_0}{1 + j\omega\tau} = |\chi|e^{j\varphi} \quad (3)$$

, where $\chi_0$ is the static susceptibility and $\omega$ is the angular frequency.

Under the mixing-frequency magnetic field, the magnetization of MNPs has a phase delay $\varphi_H$ to high-frequency field $H_H$ and a phase delay $\varphi_L$ to low-frequency field $H_L$. Based on the ferrohydrodynamic equation[30-32], as the fluid in the model dose not flow, the magnetization equation can be written as:

$$\frac{dM(t)}{t} = \frac{M_0(t) - M(t)}{\tau} \quad (4)$$

, where the equilibrium magnetization $M_0$ is followed by Langevin function[9]:

$$M_0(t) = Nm_s\left(\coth\frac{\mu_0 m_s H(t)}{k_B T} - \frac{k_B T}{\mu_0 m_s H(t)}\right) \quad (5)$$

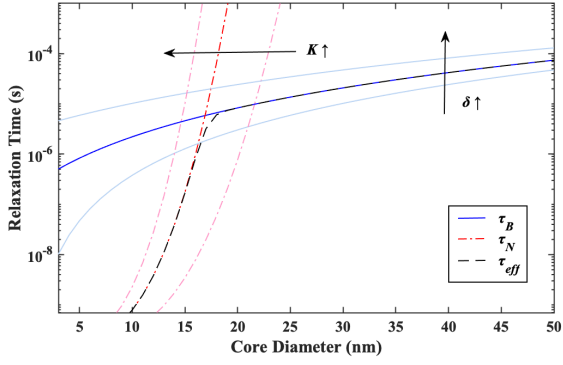

FIG. 4 Néel, Brownian and total effective relaxation time along the diameter of MNPs. Assume $T = 300K$, the thickness of coatings $\delta = 0nm, 4nm, 10nm$, the magnetic anisotropy constant $K = 10kJ \cdot m^{-3}, 20kJ \cdot m^{-3}$ and $30kJ \cdot m^{-3}, \eta = 10^{-3} Pa \cdot s$.

As $M_0(t)$ is odd to $H(t)$, which is even to $t$, the Fourier series expansion of $M_0(t)$ can be written as:
$$M_0(t) = \sum_{n=1}^{\infty} a_n \cos n\omega t \quad (6)$$
, where $a_n = \frac{\omega}{\pi} \int_{\frac{\pi}{\omega}}^{\frac{\pi}{\omega}} M_0(t) \cos n\omega t \, dt$ is the Fourier coefficient.

Take the Laplace transform of both sides of Eq.(4):
$$sL(s) - M(t=0) = -\frac{1}{\tau}[L(s) - L_0(s)] \quad (7)$$
So bring Eq. (6) to Eq. (7):
$$sL(s) - M(t=0) = -\frac{1}{\tau}\left[L(s) - \sum_{n=1}^{\infty} a_n \cos n\omega t\right] \quad (8)$$
, then:
$$L(s) = \frac{M(t=0)}{s + 1/\tau} + \sum_{n=1}^{\infty} \frac{a_n s}{(1 + s\tau)(s^2 + (n\omega)^2)} \quad (9)$$

The magnetization $M(t)$ can be solved by the inverse Laplace transform of Eq. (9):
$$M(t) = M(t=0)e^{-t/\tau}$$
$$+ \sum_{n=1}^{\infty} \frac{a_n}{1 + (n\omega\tau)^2}[\cos n\omega t + n\omega\tau \sin n\omega t - e^{-t/\tau}] \quad (10)$$

As $M(t) = 0$ and $e^{-t/\tau}$ can be ignored for too small in the measurement system, based on the conversion of trigonometric functions, Eq.(10) can be written as[18, 33]:
$$M(t) = \sum_{n=1}^{\infty} \frac{a_n}{\sqrt{1 + (n\omega\tau)^2}} \cos(n\omega t - \varphi_n) \quad (11)$$
, where $\varphi_n$ is the phase delay[19]:
$$\varphi_n = \arctan \chi''/\chi' = \arctan(n\omega\tau) \quad (12)$$

Therefore, the mixing-frequency phases at $f_H \pm 2f_L$ can be obtained as:
$$\begin{cases} \varphi(f_H + 2f_L) = \varphi_H + 2\varphi_L \\ \varphi(f_H - 2f_L) = \varphi_H - 2\varphi_L \end{cases} \quad (13)$$

According to Faraday's law, there is a $\pi/2$ phase difference between the magnetization of MNPs and the voltage signal of pick-up coils[34]. So the phases $\phi$ of the voltage signal on $f_H \pm 2f_L$ can be described as:
$$\begin{cases} \phi(f_H + 2f_L) = \varphi(f_H + 2f_L) - \frac{3\pi}{2} \\ \phi(f_H - 2f_L) = \varphi(f_H - 2f_L) - \frac{3\pi}{2} \end{cases} \quad (14)$$

Then the phase delay $\varphi_H$ can be obtained from Eq.(13) and Eq.(14) as:
$$\varphi_H = -\frac{\phi(f_H + 2f_L) + \phi(f_H - 2f_L) - 3\pi}{2} \quad (15)$$
, and the relaxation time $\tau_B$ can be calculated by:
$$\tau_B \approx \tau = \tan(\varphi_H)/2\pi f_H \quad (16)$$
Then the temperature can be determined as:
$$T = \frac{3\eta V_H}{k_B \tau_B} = \frac{A}{\tau_B} \quad (17),$$
where $A = 3\eta V_H/k_B$ is a constant almost decided by the particle size. Therefore, the temperature information can be solved according to the relaxation time of MNPs.

## B. Phase Detection of AC Magnetization

In order eliminate the remanence and obtain the pure magnetization signal of the MNPs, there are always two detection coils used to form a differential. However, it is almost impossible to achieve the same parameters of the two coils, which means the influence made by the different between the coils on the detection results cannot be ignored.

Without the MNP sample, the signals of the two detection coils A and B under the excitation field $H(t) = H_o \sin(\omega t + \phi_o)$ are:
$$\begin{cases} Sig_A = L_{A(\omega)} \cdot \omega \cdot \cos(\omega t + \phi_o + \phi_{A(\omega)}) \\ Sig_B = L_{B(\omega)} \cdot \omega \cdot \cos(\omega t + \phi_o + \phi_{B(\omega)}) \end{cases} \quad (18)$$
, where $L_{A(\omega)}$、$L_{B(\omega)}$、$\varphi_{A(\omega)}$、$\phi_{B(\omega)}$ are the effects of detection coils on signal amplitude and phase.

Therefore, the signal of the differential is:
$$Sig = Sig_A - Sig_B$$
$$= L_{A(\omega)} \cdot \omega \cdot \cos(\omega t + \phi_o + \phi_{A(\omega)})$$
$$- L_{B(\omega)} \cdot \omega \cdot \cos(\omega t + \phi_o + \phi_{B(\omega)}) \quad (19)$$

If the parameters of coils are completely the same, the remanence would be eliminated as $Sig = 0$. While it is almost impossible, the output signal of the differential amplifier is:
$$Sig_I = L_{I(\omega)} \cdot L_{A(\omega)} \cdot \omega \cdot \cos(\omega t + \phi_o + \phi_{A(\omega)} + \phi_{I(\omega)})$$
$$- L_{I(\omega)} \cdot L_{B(\omega)} \cdot \omega \cdot \cos(\omega t + \phi_o + \phi_{B(\omega)} + \phi_{I(\omega)}) \quad (20)$$
, where $\phi_{I(\omega)}$ is the phase shift produced by the differential amplifier and $L_{I(\omega)}$ is the differential amplifier magnification.

After the MNP sample put in, the magnetization coils detected is:
$$H' = H_o \sin(\omega t + \phi_o) + \chi(\omega) \cdot H_o \sin(\omega t + \phi_{s(\omega)}) \quad (21)$$
, where $\chi(\omega)$ is the MNP susceptibility and $\phi_{s(\omega)}$ is the sample phase. Then the signal of the differential is:
$$Sig' = L_{A(\omega)} \cdot \omega \cdot \cos(\omega t + \phi_o + \phi_{A(\omega)})$$
$$+ L_{A(\omega)} \cdot \chi(\omega) \cdot H_o \cos(\omega t + \phi_{s(\omega)} + \phi_{A(\omega)})$$
$$- L_{B(\omega)} \cdot \omega \cdot \cos(\omega t + \phi_o + \phi_{B(\omega)}) \quad (22)$$
, the output signal of the differential amplifier is
$$Sig'_I = Sig_I + L_{I(\omega)} \cdot L_{A(\omega)} \cdot \chi(\omega)$$
$$\cdot H_o \cos(\omega t + \phi_{s(\omega)} + \phi_{A(\omega)} + \phi_{I(\omega)}) \quad (23)$$

The difference signal before and after sample release is:
$$Sig'_I - Sig_I = L_{I(\omega)} \cdot L_{A(\omega)} \cdot \chi(\omega)$$
$$\cdot H_o \cos(\omega t + \phi_{s(\omega)} + \phi_{A(\omega)} + \phi_{I(\omega)}) \quad (24)$$

Then the phase of MNP sample can be calculated by:
$$\phi_{s(\omega)} = phase[Sig'_I - Sig_I] - phase[Sig_A] - \phi_{I(\omega)} + \phi_o \quad (25)$$

The steps of Eq.(24) and Eq.(25) are carried out in LabVIEW after the data is collected into the computer.

Therefore, under the mixing-frequency magnetic field, after the $\phi(f_H \pm 2f_L)$ and the $\varphi_H$ calculated by Eq.(25) and Eq.(15) respectively, the temperature information can be solved according to the relationship between the relaxation time and temperature.

## III. EXPERIMENT

The structure of the measurement system is shown in Fig.FIG. 5. The commercial iron oxide MNPs used for study is SHP-30 (Ocean NanoTech, USA) with 30nm core size, of which the core diameter distribution was obtained by TEM and the hydrodynamics diameter distribution was determined by nanoparticle size analyzer, as shown in Fig.FIG. 6. A USB-6356 data acquisition system (NI, USA) and an AE-7224 power amplifier (AE Techron, USA) were used to drive solenoid to generate the excitation field. The solenoid coil has a resistance of 1.29 Ohms and an inductance of 24mH. Two coils with similar characteristics were used to detect the signal. A differential amplifier was used to amplify the signal with a gain of 1000, then the signal was collected by the data acquisition system and transmitted to a computer for storage and processing. A FOB669A fiber optic thermometer (OMEGA, Norwalk CT, USA) was used to measure the standard temperature. Compared with the previous system, a signal channel is added to input the signal of the pick-up coil directly to the data acquisition system. In order to weaken the influence of the electromagnetic field in the air on the signal, the differential amplifier is put into a permalloy shielding box, and the phase shift produced by the differential amplifier with frequency is shown in Fig.FIG. 7.

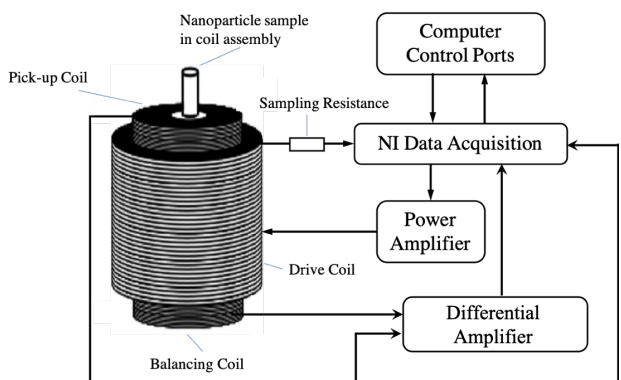

FIG. 5 Diagram of the apparatus used to measure the MNPs' signal.

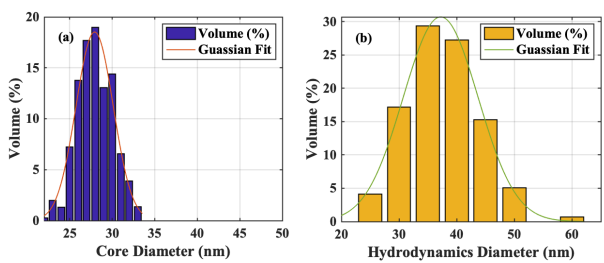

FIG. 6 Core diameter distribution (a) and Hydrodynamics diameter distribution (b) of SHP-30.

Set the high frequency $f_H = 6k$Hz and the low frequency $f_L = 1.57k$Hz to avoid the mixing frequency $f_H \pm 2f_L$ is a multiple of 50Hz, which could reduce the influence of 50Hz power frequency on test results. The amplitudes of the two magnetic fields affect the SNR of signal in different ways, which would further affect the temperature error as Fig.FIG. 2 shown. Theoretically, increasing the intensity of magnetic field can improve the SNR, while the effective relaxation time of magnetic fluid would be shortened. As the maximum power of AE-7224 is 1100W and the maximum output voltage is 180V, the magnetic field strength should be selected reasonably.

In the case of fixed $\mu_0 H_H = 0.36$mT, the spectrum simulation results of signals under different magnetic field strength ratios are shown in Fig.FIG. 8. The particle size was 30 nm and set the temperature as 300K.

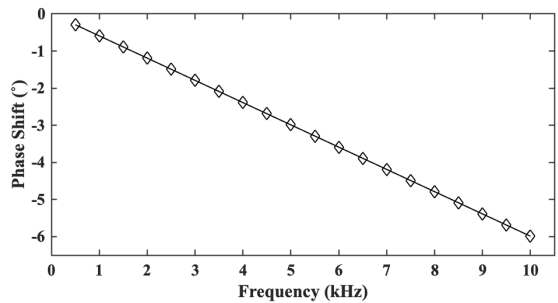

FIG. 7 The phase shift produced by the instrumentation amplifier with frequency

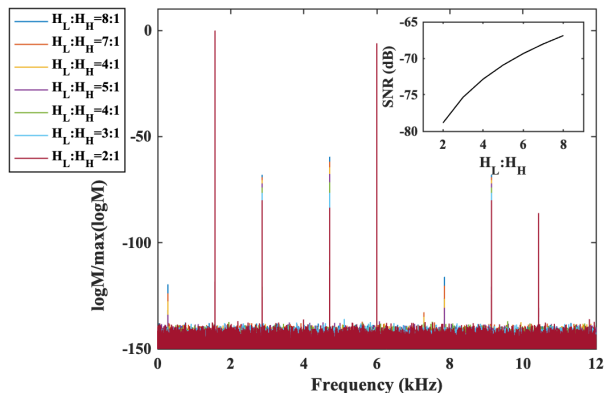

FIG. 8 Simulation of the normalized logarithmic spectrum of the magnetic nanoparticle magnetization under different magnetic field strength ratios.

## IV. RESULTS AND DISCUSSION

Based on the simulation shown as Fig.7, the excitation coil generated 6kHz AC field with 0.36mT and 1.57kHz with 1.98mT. The detection coil A has a resistance of 10.4177 Ohms and an inductance of 1.64741mH, while the coil B has a resistance of 10.6454 Ohms and an inductance of

1.70752mH. These parameters were detected by Agilent 4294A precision impedance analyzer (Keysight, USA). As hyperthermia need to raise the temperature of the tumor from 37°C to 42-45°C[24-27], set the temperature range of the experiment as 310-320K.

## A. Phase Measurement

In order eliminate the remanence and obtain the pure magnetization signal of the MNPs, there are always two detection coils used to form a differential. However, it is almost impossible to achieve the same parameters of the two coils, which means the influence made by the different between the coils on the detection results cannot be ignored.

As shown in Fig.FIG. 9, the phase at the high frequency calculated by the phases at the mixing frequencies could reduce the phase shift produced by the coil parameters changed with temperature. When the ambient temperature changed, the rate of change of the calculated phase was 0.05°/K, while the rate of change of the directly measured phase was 0.57°/K.

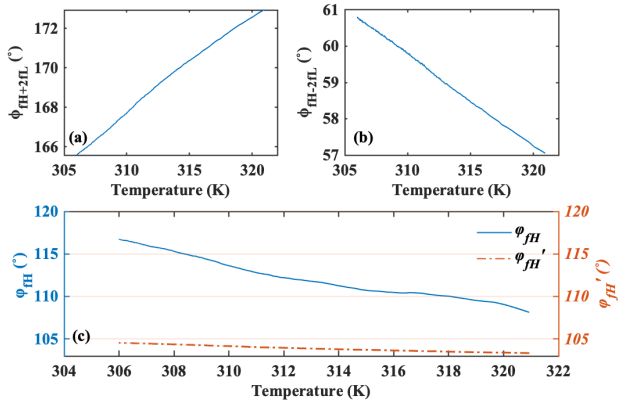

FIG. 9 Phases of mixing-frequency method. (a) and (b) are the phases at frequency $f_H \pm 2f_L$. (c) are the directly measured phase $\varphi_H$ and the calculated phase $\varphi_H'$.

The reason why the influence of temperature on the phase was not completely eliminated may be that the resistance value of the coils was affected by the temperature.

## B. Temperature Measurement

In the experiment, the static temperature of the sample and the dynamic temperature of the cooling process are measured respectively. The temperature measurement results are shown in Fig.FIG. 10 and Fig.FIG. 11, where ET is the estimated temperature and TT is the standard temperature. The temperature error reflects the difference between the estimated temperature and the standard temperature.

In order to maintain the sample temperature stable during the static temperature measurement, the method of water bath temperature control was used to control the sample temperature. Fig.FIG. 10(a) shows the temperature error for two minutes when the sample temperature was 315.6K. The maximum temperature error was 0.067K and the standard deviation was 0.0267K. Fig.FIG. 10(b) shows the temperature error in the range 310-320K, in which it can be seen that the error was less than 0.1K.

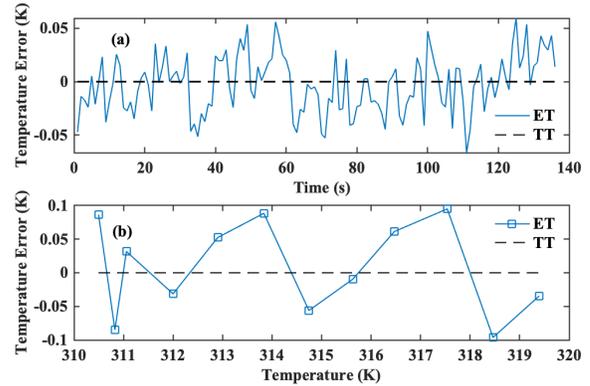

FIG. 10 Static temperature measurement results obtained using the mixing-frequency model. (a) is the temperature error of one temperature point, (b) is temperature errors of multi-temperature points in the range from 310K to 320K.

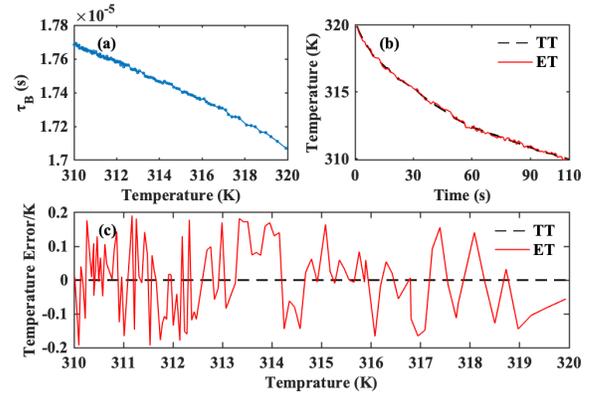

FIG. 11 Dynamic temperature measurement results obtained using the mixing-frequency model. (a) shows the relaxation time changed with temperature, (b) is a comparison of the estimated temperature (ET) with the standard temperature (TT), (c) is the temperature error curve.

In the cooling process, the sample temperature dropped from 320K to 310K. Fig.FIG. 11(a) shows the relaxation time of the magnetic nanoparticles as a function of the standard temperature. It can be seen from Fig.FIG. 11(b) that the temperature estimates obtained using this method matched the standard temperature well. Fig.FIG. 11(c) is the temperature error curve, and it can be seen that the error was less than 0.2K, and the standard deviation was 0.107k.

The main reason, why the dynamic temperature measurement error was worser than the static temperature measurement error, is that although the method could weaken the influence of the temperature on the detection coil parameters, it cannot completely eliminate the influence, as shown in Fig.FIG. 9.

## IV. RESULTS AND DISCUSSION

In summary, a mixing-frequency method has been used to measure the relaxation time and temperature of MNPs. The static temperature measurement error is less than 0.1k and the dynamic temperature measurement error is less than 0.2K.

In this method, two sinusoidal magnetic fields were applied simultaneously to make MNPs generate a magnetization signal. The phase of the $f_H \pm 2f_L$ harmonics were used to obtain the phase information at $f_H$. This phase solution method has been proved to be able to reduce the phase shift caused by the change of coil parameters with temperature. Considering the characters of two detection coils may not be completely consistent, a phase detection method which can eliminate the phase deviation introduced by the parameter difference of detection coils is proposed.

However, the phase detection method still requires the selection of detection coil in advance. If the parameter difference between the two coils is too large, the output signal of the differential amplifier will exceed the input upper limit of the data acquisition system, and then the next steps cannot be carried out. The entire parts of the measurement hardware system also affect the measurement results, which would be studied in the following research.

## ACKNOWLEDGMENTS

This work was supported by the project of the project of 61973132(NSFC).